\documentclass[%
 reprint,amsmath,amssymb,aps,nofootinbib,bibtex]{revtex4-1}
\usepackage{graphicx}% Include figure files
\usepackage{dcolumn}% Align table columns on decimal point
\usepackage{bm}% bold math
\usepackage{overpic}

\def\balpha{\mbox{\boldmath $\alpha$}}

\def\btau{\mbox{\boldmath $\tau$}}
\def\bnu{\mbox{\boldmath $\nu$}}

\def\bn{\mbox{\bf n}}
\def\bt{\mbox{\bf t}}
\def\bc{\mbox{\bf c}}
\def\bu{\mbox{\bf u}}
\def\bb{\mbox{\bf b}}
\def\bs{\mbox{\bf s}}
\def\bd{\mbox{\bf d}}

\def\Im{\mathop{\rm Im}\nolimits}

\def\frac#1#2{{{#1}\over{#2}}}
\def\tfrac#1#2{{\textstyle{{#1}\over{#2}}}}

\def\half{\tfrac{1}{2}}

\begin{document}

%\preprint{APS/123-QED}

\title{On the Corner Elements of the CKM and PMNS Matrices}\thanks{Work supported in part by Spanish MICINN and FEDER (EC) under grant FPA2011-23596 and Generalitat Valenciana under grant GVPROMETEO2010-056.}

\author{Michael J BAKER}\email{michael.baker@uv.es} 
\affiliation{%
Departament Fisica Teorica and IFIC, Centro Mixto CSIC, Universitat de Valencia,
Calle Dr. Moliner 50, E-46100 Burjassot (Valencia), Spain}

\author{Jos\'e BORDES}\email{jose.m.bordes\,@\,uv.es} 
\affiliation{%
Departament Fisica Teorica and IFIC, Centro Mixto CSIC, Universitat de Valencia,
Calle Dr. Moliner 50, E-46100 Burjassot (Valencia), Spain}

\author{CHAN Hong-Mo}%
\email{h.m.chan\,@\,stfc.ac.uk}
\affiliation{%
Rutherford Appleton Laboratory,\\
Chilton, Didcot, Oxon, OX11 0QX, United Kingdom
}%

\author{TSOU Sheung Tsun}%
\email{tsou\,@\,maths.ox.ac.uk}
\affiliation{%
Mathematical Institute, University of Oxford,\\
24-29 St. Giles', Oxford, OX1 3LB, United Kingdom
}%
\date{\today}% 

\begin{abstract}

Recent experiments show that the top-right corner element ($U_{e3}$) 
of the PMNS, like that ($V_{ub}$) of the CKM, matrix is small but 
nonzero, and suggest further via unitarity that it is smaller than 
the bottom-left corner element ($U_{\tau 1}$), again as in the CKM 
case ($V_{ub} < V_{td}$).  An attempt in explaining these facts 
would seem an excellent test for any model of the mixing phenomenon.  
Here, it is shown that if to the assumption of a universal rank-one 
mass matrix, long favoured by phenomenologists, one adds that this 
matrix rotates with scale, then it follows that (A) by inputting 
the mass ratios $m_c/m_t, m_s/m_b, m_\mu/m_\tau$, and $m_2/m_3$, 
(i) the corner elements are small but nonzero, (ii) $V_{ub} < 
V_{td}$, $U_{e 3} < U_{\tau 1}$, (iii) estimates result for the 
ratios $V_{ub}/V_{td}$ and $U_{e 3}/U_{\tau 1}$, and (B) by 
inputting further the experimental values of $V_{us}, V_{tb}$ and 
$U_{e2},U_{\mu 3}$, (iv) estimates result for the values of the 
corner elements themselves.  All the inequalities and estimates 
obtained are consistent with present data to within expectation for the approximations made.

\end{abstract}
% insert suggested PACS numbers in braces on next line

\maketitle

The purpose of this note is to try to understand the smallness
of the corner elements of the quark \cite{cabibbo,km} and 
lepton \cite{ponte,mns} mixing matrices 
and their asymmetry about the diagonals.  It is shown that a 
scheme with a rank-one rotating mass matrix (R2M2) devised to 
explain the hierarchical masses and mixing of fermions 
\cite{r2m2} will automatically also give, as a bonus, the said 
asymmetry correctly.

To theoreticians, the mixing matrices of quarks and leptons are
a bit of an embarrassment.  While their experimental colleagues
have improved the measurements of the CKM  
matrix elements to
an impressive accuracy and gained an increasingly clear picture 
even of the PMNS 
matrix, which is incredibly difficult to acquire, no
commonly accepted theoretical understanding has been achieved
even of the qualitative features, let alone a calculation of
the matrix elements to the accuracy now measured in experiment.

The latest injection from experiment is a batch of impressive new 
results, in order of appearance \cite{T2K,Minos,DChooz,DayaBay,
RENO}, which give a nonzero value for the upper corner element 
usually called $U_{e3}$ of the PMNS matrix.  Now this $U_{e3}$
was widely expected to vanish based on some symmetry arguments
which treated the mixing of leptons differently from that of
quarks \cite{Scott}.  The new measured value for this, however,
is in fact much larger in magnitude than the corresponding element
$V_{ub}$ in the CKM matrix.  Indeed, a combined fit by Blondel
\cite{Blondel} of the first 4 experiments gives a value of:
\begin{equation}
\sin^2 (2 \theta_{13}) = 0.084 \pm 0.014.
\label{blondel}
\end{equation}
If we take this central value of $\theta_{13}$ and assume, as some 
of these experiments do, that $\sin^2 (2 \theta_{23}) = 1$ and the 
CP phase $\delta = 0$, we get the PMNS matrix as:
\begin{equation}
U_{\rm PMNS} = \left( \begin{array}{lll} 0.820 & 0.554 & 0.146 \\
                                      0.482 & 0.528 & 0.699 \\
                                      0.310 & 0.644 & 0.699
                  \end{array} \right),
\label{UPMNS}
\end{equation}
as compared to the measured CKM matrix given in \cite{databook}:
\begin{equation}
V_{\rm CKM} = \left( \begin{array}{ccc} 
                             0.97428 & 0.2253 & 0.00347 \\
                             0.2252 & 0.97345 & 0.0410 \\
                             0.00862 & 0.0403 & 0.999152
                                     \end{array} \right),
\label{VCKM}
\end{equation}
where in all cases the central values of the matrix elements 
are given.   For these we quote 
only their absolute values, as we shall deal mainly with these, only
returning to the 
important question of the Kobayashi--Maskawa CP-violating phase
\cite{mns} at the end.  

One notes that the two matrices (2) and (3) share some
qualitative features.  In both, the corner elements are
rather small, and have the same sign of asymmetry about 
the diagonal.  
(It is enough to look at the corner elements since the asymmetry 
in the other elements would then follow by unitarity, the norms 
of all rows and columns being unity).  In both matrices, the 
bottom-left corner element is larger than the top-right corner 
element by about a factor of 2.  This asymmetry, together with
the small values of the corner elements, are what particularly 
interest us here, for the prediction of both would be a delicate 
test no {\it ad hoc} model is likely to reproduce. 

The similarity between the two matrices suggests, to us at least,
that they be treated similarly and understood together as two 
facets of the same phenomenon, and the R2M2 scheme is an attempt 
to do so.  We will give first a very brief outline of this scheme
to facilitate future discussions.  For details, 
the reader is referred to \cite{r2m2}, a recent review.

The R2M2 scheme was suggested some years \cite{physcons,ckm,phenodsm} ago as a possible explanation 
for the hierarchical mass spectrum and mixing pattern
of quarks and leptons observed in experiment and incorporated 
{\it per se} into the standard model.  We note first that any 
fermion mass matrix can, by a judicious relabelling of the $su(2)$ 
singlet right-handed fields, be cast into a form with no dependence 
on $\gamma_5$ \cite{Weinbergren} so that any rank-one mass matrix can 
be written without loss of generality in the form
\begin{equation}
m = m_T {\balpha}{\balpha}^{\dagger},
\label{mfact}
\end{equation}
in terms of a unit vector $\balpha$ in generation space.  
This means that there is only one massive generation.  With 
$\balpha$ ``universal'' in 
the sense of being independent of the fermion types $T$ (i.e., 
whether up or down, or whether leptons or quarks), it further implies
that there is no mixing between up and down states.  
Now such
a form of the fermion mass matrix 
has long been suggested by phenomenologists \cite{Fritsch,Harari} as 
a good starting point for understanding mass hierarchy and mixing.
What R2M2 adds to this is that the vector $\balpha$ rotates
with changing scales under renormalization.

How R2M2 can lead to mixing and mass hierarchy can be seen most simply
by considering just the two heaviest generations, and assuming 
for further simplification that ${\balpha}$ is real and that $m_T$ is
scale-independent.  The eigenvector $\balpha$ at scale $\mu = m_t$ is the
state vector \bt\ of $t$, and the orthogonal vector \bc\ is that of $c$.
As the scale decreases,  $\balpha$ rotates through an angle, say $
\theta_{tb}$, when it reaches the scale $\mu = m_b$, where it becomes
the state vector \bb\ of $b$.  The vector orthogonal to \bb\ is then the
state vector \bs\ of $s$.  We have therefore two dyads, $\{{\bf t}, {\bf
c}\}$ and $\{{\bf b}, {\bf s}\}$, linked by the non-identity mixing matrix
\begin{equation}
\left( \begin{array}{cc} V_{cs} & V_{cb} \\ V_{ts} & V_{tb} \end{array}
   \right) = \left( \begin{array}{cc}  {\bf c} \cdot{\bf s}  
                             &  {\bf c} \cdot{\bf b}  \\
                                {\bf t} \cdot{\bf s} 
                             &  {\bf t} \cdot{\bf b}
             \end{array} \right )
           = \left( \begin{array}{cc} \cos \theta_{tb} & -\sin \theta_{tb} \\  
                \sin \theta_{tb} & \cos \theta_{tb} \end{array} \right),    
\label{UDmix2}
\end{equation}
which is the 2-generation analogue of the CKM matrix.

As to hierarchical masses, denoting by $m_U$ and $m_D$ the values of
$m_T$ for $U$ and $D$ type quarks, we have $m_t = m_U$ and $m_b =
m_D$.  At $\mu = m_t$ the eigenvector \bc\ has zero eigenvalue, but
this is not the mass of the $c$ state, which should be evaluated at 
$\mu = m_c$.  Indeed, $m_c$ is to be taken as the 
solution to the equation 
\begin{equation}
\mu = \langle {\bf c}|m(\mu)|{\bf c} \rangle 
    = m_U |\langle {\bf c}|{\balpha}(\mu) \rangle|^2.
\label{solvmc}
\end{equation}
A nonzero solution exists since the scale on the LHS decreases from 
$\mu=m_t$ while the RHS increases from zero at that scale.  At $\mu <
m_t$ the vector will have rotated from \bt\ to a different direction
so that it will have acquired a nonzero component,  say 
$\sin \theta _{tc}$, in the direction of ${\bf c}$ giving
\begin{equation}
m_c = m_t \sin^2 \theta_{tc}.
\label{mc2}
\end{equation}
Further $m_c$ will be small if the rotation is not too fast.
Similarly for $m_s$.

An interesting point to note here is that 
the mass matrix (\ref{mfact}) remains rank-one throughout.  Yet,
simply because the mass matrix rotates, the lower generation $c$
acquires a nonzero mass, as if by ``leakage'' from the heavy
state $t$.  This ``leakage mechanism'' is a very special property 
of R2M2 to which we shall have occasion to return.    

Basically the same arguments apply to the realistic case when
all 3 generations are taken into account,  although 
the analysis  becomes a little more complicated.  As the 
scale changes, the unit vector $\balpha$ now traces out a curve 
on the unit sphere, which can bend in
two directions, either along the sphere or sideways, and it 
can also twist, and it is these contortions of the rotation
trajectory as the scale changes which will now determine the
fermion mass and mixing patterns.  Nevertheless, applying
exactly the same physical arguments as before, one deduces,
say for the $U$-type quarks, the following formulae
\begin{eqnarray}
{\bf t} & = & {\balpha}(m_t), \nonumber \\
{\bf c} & = & {\bf u} \times {\bf t}, \nonumber \\
{\bf u} & = & \frac{{\balpha}(m_c) \times {\balpha}(m_t)}
   {|{\balpha}(m_c) \times {\balpha}(m_t)|},
\label{tcutriad}
\end{eqnarray}
and
\begin{eqnarray}
m_t & = & m_U, \nonumber \\
m_c & = & m_U |{\balpha} (m_c) \cdot{\bf c}|^2, \nonumber \\
m_u & = & m_U |{\balpha} (m_u) \cdot{\bf u}|^2.
\label{hiermass}
\end{eqnarray}

Together, these 2 sets of coupled equations allow us to
evaluate both the state vectors and the masses.  Similar
equations and remarks apply also to $D$-type quarks as
well as to the leptons.  With the state
vectors so determined, the mixing matrices can then be 
evaluated, e.g., for quarks:
\begin{equation}
V_{\rm CKM} \sim \left( \begin{array}{ccc}
   {\bf u} \cdot{\bf d}  &  {\bf u} \cdot
{\bf s}  &  {\bf u} \cdot{\bf b}  \\
    {\bf c} \cdot{\bf d}  &  {\bf c} \cdot
{\bf s}  &  {\bf c} \cdot{\bf b}  \\
    {\bf t} \cdot{\bf d}  &  {\bf t} \cdot
{\bf s}  &  {\bf t} \cdot{\bf b}  
          \end{array} \right).
\label{Vckm}
\end{equation}
The expression for the lepton mixing matrix $U_{\rm PMNS}$ would
be similar.

\begin{figure*}
\centering
%\begin{overpic}[width=12cm]{rotacurvefitx.pdf}
\begin{overpic}[width=12cm]{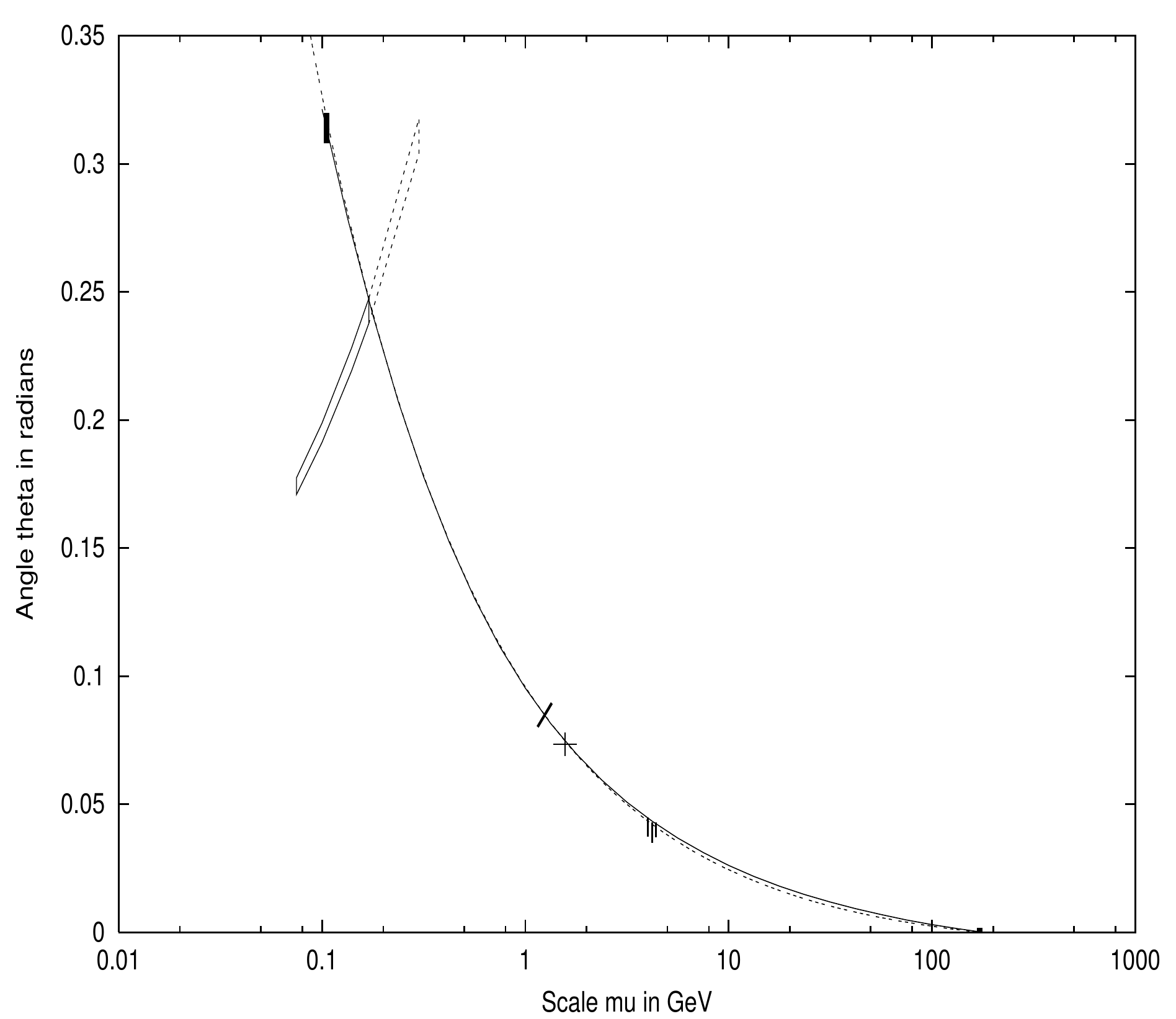}
\put(15,75){$m_{\mu}/m_{\tau}$}
\put(35,60){$m_s/m_b$}
\put(48,28){$m_c/m_t$}
\put(51,24){$\tau$}
\put(57,17){$V_{tb},V_{ts},V_{cb}$}
\put(80,10){$t$}
\end{overpic}
\caption{The planar rotation curve taken from \cite{cevidsm} for illustration.
The solid curve shown was the best (exponential) fit to the
data, which have not changed much since then, except for 
the point $m_s/m_b$ which figures little in the fit because 
of its large errors.  The dotted curve was a fit with a
phenomenological (but theoretically incomplete) model then 
being worked on, which is by now largely superseded.}
\label{planarplot}
\end{figure*}
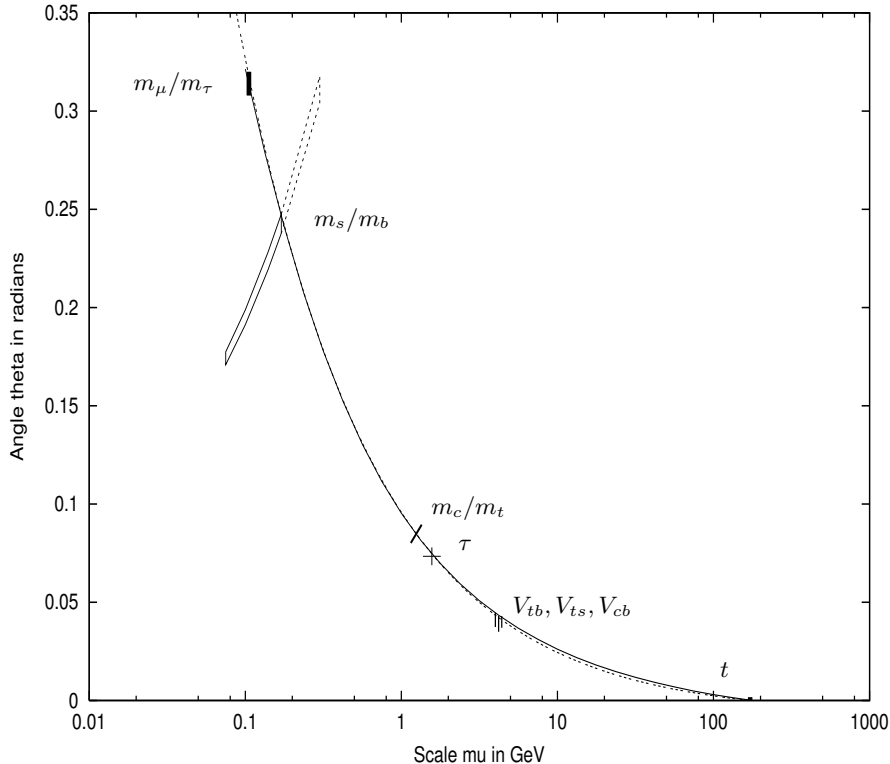

An unusual outcome of the R2M2 hypothesis, as outlined 
above, is in giving nonzero masses to all fermions while 
the fermion mass matrix itself remains of rank one and 
chiral invariant throughtout.  At first sight, this may
seem counter-intuitive, due to the unfamiliarity of the 
new conditions introduced by a rotating mass matrix, which 
require a revision of some common notions deduced earlier
from mass matrices which do not rotate.  For a discussion 
of these points which, though straightforward, need some 
patience to go through, the reader is referred elsewhere, 
e.g. to \cite{r2m2}, especially section 1.4 therein.

That the mass spectra and mixing matrices so obtained from a
rank-one rotating mass matrix (R2M2) actually do resemble those
observed in experiment can be checked in two ways: (i) invent a
model trajectory, then evaluate in the manner indicated the mass
spectra and mixing matrices of quarks and leptons to compare 
with experiment, or conversely, (ii) fit a trajectory through 
the experimental data on these quantities.  Both have been tried
with encouraging success \cite{r2m2,btfit,phenodsm}.  We quote here in
Figure \ref{planarplot} a particularly simple example from an
early work \cite{cevidsm} which will be of use to us later.  As
seen in this figure, the rotation angle $\theta$ obtained via 
(\ref{UDmix2}) and (\ref{mc2}) from the experimental masses and
mixing angles in the 2-generation simplification all fall on a
smooth curve as a function of $\mu$ as expected. 

However, both the above methods (i) and (ii) for testing the R2M2 
hypothesis involve some manipulations of the data or assumptions
about the shape of the rotation trajectory, and may in the process
obscure somewhat the basic simplicity of the result.  What we wish
to do here, in contrast, is to so drastically 
simplify the argument as to make the correlations between the 
various bits of data immediately obvious, in particular those
in relation to the corner elements of the mixing matrices.  To
do so, we rely on the fact that most of the rotation angles
involved are relatively small so that use can be made \cite{features} of the 
differential Serret--Frenet--Darboux formulae for curves lying on a
surface \cite{Dorcarmo}.

At every point of the trajectory for $\balpha$, we set up 
a Darboux triad, consisting of first the normal to the surface \bn,
then the tangent vector $\btau$ to the trajectory, and thirdly  
the normal $\bnu$ to both the above, normalized and orthogonal 
to one another.  The Serret--Frenet--Darboux formulae then say
\begin{eqnarray}
\bn'=\bn(s+\delta s) & =& \bn(s) - \kappa_n \btau (s)\,\delta s 
+ \tau_g \bnu(s)\,\delta s,  \nonumber \\
\btau'=\btau(s+\delta s) &=& \btau (s) + \kappa_n\bn(s)\,\delta s
+ \kappa_g \bnu (s)\,\delta s,  \nonumber \\
\bnu' = \bnu(s+\delta s) &=& \bnu(s) - \kappa_g \btau (s)\,\delta s -\tau_g
\bn(s)\,\delta s,
\label{SFD1}
\end{eqnarray}
to first order in $\delta s$, a small increment in the arc-length
$s$, where $\kappa_n$ is known as the normal curvature (bending
along the surface), $\kappa_g$ the geodesic curvature (bending
sideways) and $\tau_g$ the geodesic torsion (twist).  For the 
special case here of a curve on the unit sphere, the surface normal \bn\ is
the radial vector \balpha, $\kappa_n = 1$
and $\tau_g = 0$, so that the formulae reduce to
\begin{eqnarray}
\balpha'=\balpha(s+\delta s) & =& \balpha(s) - \btau (s)\,\delta s, 
\nonumber \\
\btau'=\btau(s+\delta s) &=& \btau (s) + \balpha(s)\,\delta s
+ \kappa_g \bnu (s)\,\delta s,  \nonumber \\
\bnu' = \bnu(s+\delta s) &=& \bnu(s) - \kappa_g \btau (s)\,\delta s.
\label{SFD2}
\end{eqnarray} 

At $\mu = m_t$, we recall that $\balpha$ coincides with the
state vector ${\bf t}$ for the $t$ quark.  As the three quarks $t$,
$c$ and $u$ are by definition independent quantum states, the
state vectors ${\bf c}$ and ${\bf u}$ must be orthogonal 
to $\balpha$ and to each other.  They must therefore be related
to $\btau$ and $\bnu$  by a rotation about $\balpha$ by an 
angle, say $\omega_U$,
\begin{equation}
({\bf t}, {\bf c}, {\bf u}) = (\balpha, \cos \omega_U \btau
+ \sin \omega_U \bnu, \cos \omega_U \bnu - \sin \omega_U \btau).
\label{Utriad}
\end{equation}
Similarly, for the $D$-type quarks we can write
\begin{equation}
({\bf b}, {\bf s}, {\bf d}) = (\balpha', \cos \omega_D \btau'
+\sin \omega_D \bnu', \cos \omega_D \bnu' - \sin \omega_D \btau'),
\label{Dtriad}
\end{equation}
where $\{\balpha', \btau', \bnu'\}$ is the Darboux triad taken at
$\mu = m_b$, and $ \omega_D$ is the corresponding rotation angle about
$\balpha'$.

Now the angle $\theta_{tb}$ between ${\bf t}$ and ${\bf b}$,
which on the unit sphere is also the arc-length between the 
two, is envisaged in the rotation scheme to be rather small.
As seen in Figure \ref{planarplot},
the rotation angle deduced from data seems to approach 
an asymptote at $\mu = 
\infty$, the best fit to the data there being in fact
an exponential.  The rotation will speed up as $\mu$ decreases, but
for $\mu$ between $m_t$ and $m_b$, the rotation remains rather small.  
Indeed, according to (\ref{UDmix2}) it is given by the CKM 
matrix element $V_{tb} = \cos \theta_{tb}$.  Experimentally, 
as seen above in (\ref{VCKM}), this is measured to have the 
value
\begin{equation}
V_{tb} = 0.999152^{+0.000030}_{-0.000045},
\label{Vtbexpt}
\end{equation}
which gives
\begin{equation}
\theta_{tb} = \delta s \sim 0.04119^{+0.000166}_{-0.000075}.
\label{thetatb}
\end{equation}
To this order of smallness then, we can take the expressions 
in (\ref{SFD2}) as the Darboux triad at $\mu = m_b$ and hence 
(\ref{Dtriad}) as the $D$-triad.  This gives immediately the 
CKM matrix, according to (\ref{Vckm}), as
\begin{widetext}
\begin{equation}
V_{\rm CKM} = \left( \begin{array}{ccc} \cos (\omega) - \kappa_g \sin (\omega)
    \,\theta_{tb} &  \sin (\omega) + \kappa_g \cos (\omega) \,\theta_{tb} & 
     \sin (\omega_U) \,\theta_{tb} \\ -\sin (\omega) - \kappa_g \cos (\omega) 
    \,\theta_{tb} & \cos (\omega) - \kappa_g \sin (\omega) \,\theta_{tb} &
    -\cos (\omega_U) \,\theta_{tb} \\ -\sin (\omega_D) \,\theta_{tb} & 
    \cos (\omega_D) \,\theta_{tb} & 1 \end{array} \right)
\label{Vckm2}
\end{equation}
\end{widetext}
with $\omega = \omega_D - \omega_U$.

Although correct only to first order in $\theta_{tb}$, (\ref{Vckm2}) 
exhibits succinctly some of the special 
properties arising from rotation with clear correspondence with 
experiment, which we shall now examine.  We shall do so step-by-step starting
with the least inputs and assumptions to get the general patterns,
and then proceeding to more inputs and assumptions to get actual estimates. 

First, on 
the immediate level, simply by virtue of the fact that $\theta_{tb}$ 
is small, we note already that (i) the off-diagonal elements in the 
last row and the last column are all of order $\theta_{tb}$ and 
therefore small compared to the others, (ii) the three diagonal 
elements are markedly different, the first two being equal to first 
order in $\theta_{tb}$, and both differing from unity by an amount 
of the same order, while the last stands alone, differing from unity 
by only order $\theta_{tb}^2$, (iii) the elements $V_{us}$ and 
$V_{cd}$ are equal also to first order in $\theta_{tb}$.  A glance 
at (\ref{VCKM}) shows that these are all in agreement with what is 
experimentally observed.       
 
Next, focussing now on the corner elements:
\begin{eqnarray}
V_{ub} & = & {\bf u}\cdot{\bf b} =  \sin (\omega_U) \,\theta_{tb}, \nonumber \\
V_{td} & = & {\bf t}\cdot{\bf d} =- \sin (\omega_D) \,\theta_{tb},
\label{cornerels}
\end{eqnarray}
we recall that $\omega_U$ is the angle between the two vectors 
${\bf c}$ and $\btau$ on the plane orthogonal to $\balpha = \bt$, 
where $\btau$ is the tangent to the trajectory at $\mu = m_t$ and
${\bf c}$, by (\ref{tcutriad}) above, is the vector lying on the
plane containing the vectors $\balpha(\mu=m_t)$ and $\balpha
(\mu=m_c)$.  In other words, the angle $\omega_U$ arises as a
consequence of the rotation of the vector $\balpha(\mu)$ as $\mu$
changes from $m_t$ to $m_c$, and is thus generically of the same order of
magnitude as $\theta_{tc}$ which, according to (\ref{mc2}) or 
(\ref{hiermass}) above, gives rise to the mass ratio $m_c/m_t$.  
The same conclusion applies to $\omega_D$, namely that it should 
be of the same order as $\theta_{bs}$ which gives rise to the 
mass ratio $m_s/m_b$.  Hence it follows that the two corner elements
must both be particularly small compared with the others, since
they are respectively of order $\theta_{tb}\, \theta_{tc}$ 
and $\theta_{tb}\, \theta_{bs}$ where both the angles in each of 
the products are small as a result of the rotation scenario.    
These corner elements are small basically because they are given 
by the twist of the trajectory, and the geodesic torsion $\tau_g$ 
being zero on a sphere, the twist can only arise as a second order 
effect of the rotation.  We have thereby a ready explanation in 
the rotation scheme for why the corner elements in the CKM matrix 
are so particularly small, as experimentally observed.

As a corollary of both $\omega_U$ and $\omega_D$ being small, it
follows from (\ref{Vckm2}) that the elements $V_{cb}$ and $V_{ts}$ will have
about the same value as $\theta_{tb}$, i.e., by (\ref{thetatb})
$\sim 0.041$, again as experimentally observed in (\ref{VCKM}).  That estimates can be made on these two elements so immediately, 
but not on the two similarly-placed elements $V_{us}, V_{cd}$, 
comes about in the rotation scenario because the first pair is 
proportional to the normal curvature $\kappa_n$ in the original 
Serret--Frenet--Darboux formulae (\ref{SFD1}), and $\kappa_n = 1$ 
on the unit sphere when applied to R2M2 in (\ref{SFD2}), while 
both elements of the second pair are proportional to the geodesic 
curvature $\kappa_g$ which can have any value, depending on the rotation trajectory, even on the unit sphere.

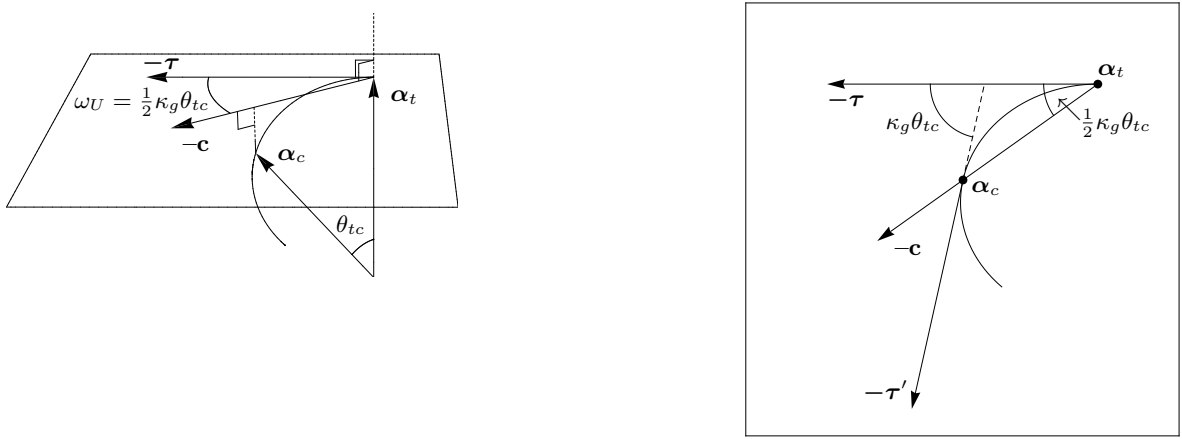
\begin{figure*}
\begin{center}
\begin{tabular}{m{6.5cm} m{6.5cm} }
\hspace{-2.6cm}
%\begin{overpic}[width=6cm]{kg17sp073d.pdf}
\begin{overpic}[width=6cm]{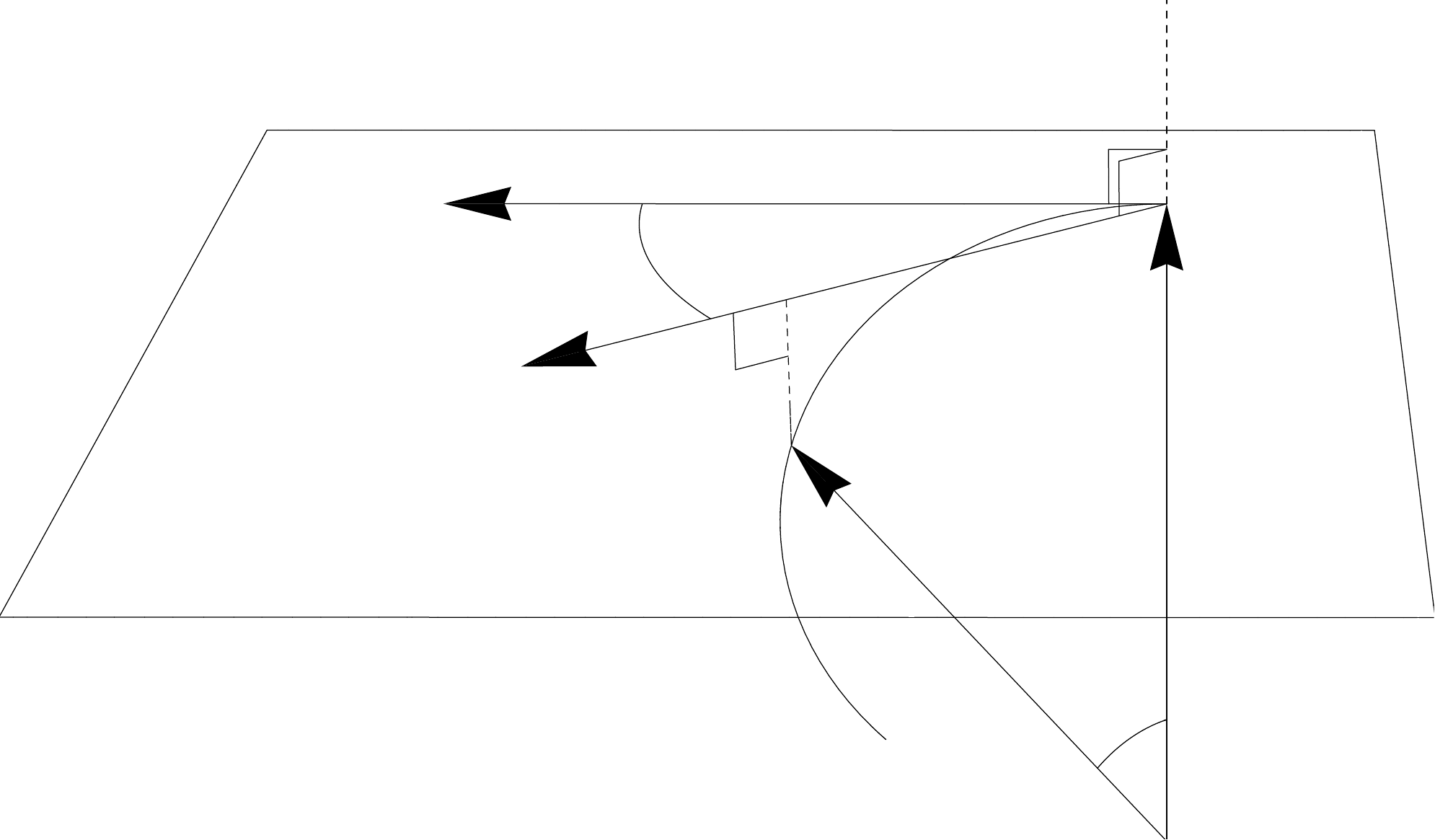}
\put(15,37.5){$\omega_U=\frac{1}{2}\kappa_g \theta_{tc}$}
\put(30,46){$\boldsymbol{-\tau}$}
\put(39,28){\textbf{--c}}
\put(60,26){$\boldsymbol{\alpha}_c$}
\put(85,39){$\boldsymbol{\alpha}_t$}
\put(73,10){$\theta_{tc}$}
\end{overpic}
\vspace{2cm}
&
\hspace{+.5cm}
%\begin{overpic}[width=6cm]{kg17sp07.pdf}
\begin{overpic}[width=6cm]{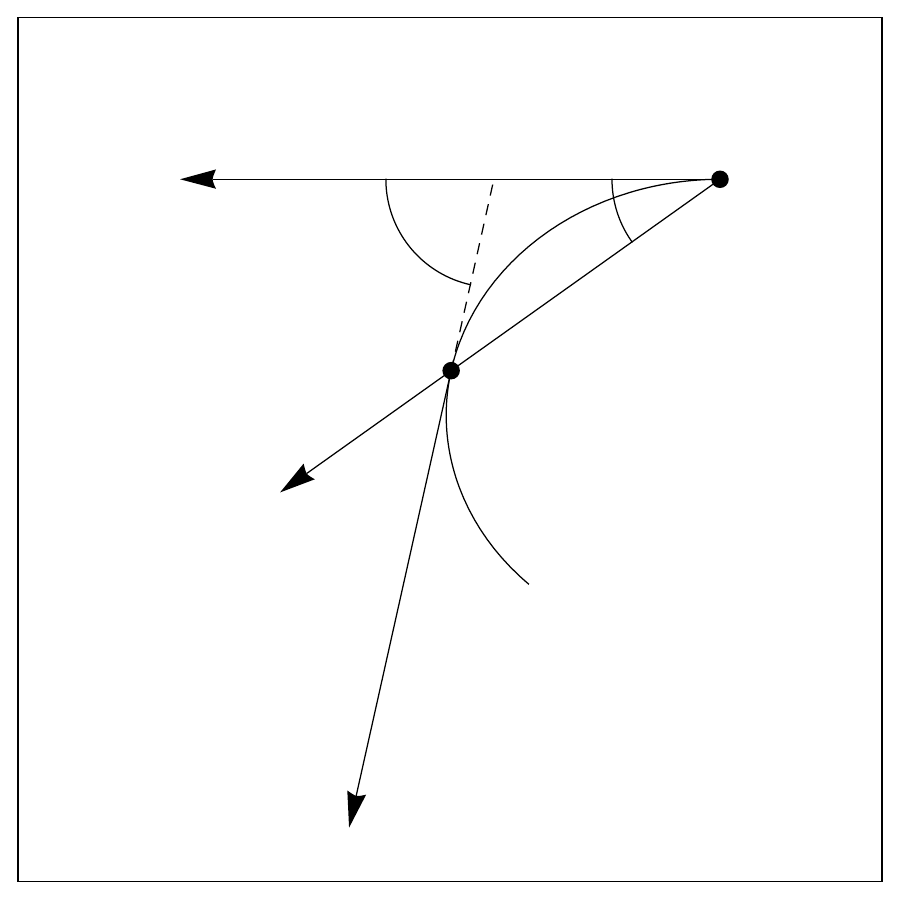}
\put(20,75){$\boldsymbol{-\tau}$}
\put(28,10){$\boldsymbol{-\tau}^\prime$}
\put(35,42){\textbf{--c}}
\put(33,70){$\kappa_g\theta_{tc}$}
\put(71,73){$\nwarrow$}
\put(76,70){$\frac{1}{2}\kappa_g\theta_{tc}$}
\put(80,82){$\balpha_t$}
\put(52,56){$\balpha_c$}
\end{overpic}
\end{tabular}
\end{center}
\caption{Diagrams explaining the relation between $\kappa_g$ and
the angle $\omega_U$:  (a) how it looks in space;  (b) projection
of the relevant part into the tangent plane at $t$ orthogonal to 
$\balpha(\mu=m_t)$.  Here we write $\balpha_t$ for $\balpha(\mu=m_t)$ and $\balpha_c$ for $\balpha(\mu=m_c)$.}
\label{kappagomega}
\end{figure*}

Going further, we notice in Figure \ref{planarplot} that rotation
seems to be speeding up from $\mu \sim m_t$ to $\mu \sim m_b$.
More concretely, taking the mass values in GeV,
\begin{eqnarray}
m_t& =& 172.9 \pm 0.6 \pm 0.9\nonumber \\
m_b &=& 4.19^{+0.18}_{-0.06}\nonumber\\
m_c &=& 1.29^{+0.05}_{-0.11}\nonumber\\ 
m_s &=& 0.100^{+0.030}_{-0.020} 
\label{mtbcs}
\end{eqnarray}

cited by PDG \cite{databook}, one obtains,
\begin{eqnarray}
\theta_{tc} &\sim& \sqrt{m_c/m_t} \ \sim \ 0.086^{+0.0019}_{-0.0039},\nonumber\\ 
\theta_{bs} &\sim& \sqrt{m_s/m_b} \ \sim \ 0.154^{+0.023}_{-0.019}.
\label{thetatcbs}
\end{eqnarray}
Now in the rotation picture, as explained above,  
$\omega_U$ is closely related to
$\theta_{tc}$ and $\omega_D$ to $\theta_{bs}$, so that
in as much as $\theta_{tc} < \theta_{bs}$, so will 
\begin{equation}
V_{ub} < V_{td},  
\label{Vub<Vtd}
\end{equation}
which is the correct asymmetry in the corner elements observed in 
experiment, again readily explained here by rotation.  

As a corollary of (\ref{Vub<Vtd}), one can deduce from
(\ref{Vckm2}) that
\begin{equation}
V_{cb} > V_{ts},
\label{Vcb>Vts}
\end{equation}
although this would also follow from unitarity, but the following
inequalities
\begin{equation}
\frac{V_{ub}}{V_{td}} < \frac{V_{ts}}{V_{cb}} < \frac{V_{cd}}{V_{us}},
\label{3ineq}
\end{equation}
implied as well by (\ref{Vub<Vtd}) and (\ref{Vckm2}) are not so 
obvious, and are equally satisfied by experiment.

One can go further still to make a semi-quantitative estimate for 
the size of the asymmetry between the two corner elements as follows.  
The angles $\theta_{tc}$ and $\theta_{bs}$ are still fairly small, 
to which one can reasonably apply again the Serret--Ferret--Darboux 
formulae of (\ref{SFD2}).  In particular, the second equation there 
for $\btau(s + \delta s)$, giving the change in direction of the 
tangent vector $\btau$ in terms of the rotation angle $\delta s$,
resolves this change into two components, one along the radius of the sphere 
which is proportional to the normal curvature $\kappa_n$, the 
other ``sideways'' on the tangent plane orthogonal to $\balpha$ 
which is proportional to the geodesic curvature $\kappa_g$, as 
illustrated in Figure \ref{kappagomega}(a).

What interests us here is the second component proportional to 
$\kappa_g$ which can be thought of as the change in direction 
of the tangent to the rotation curve, but now projected into 
the plane orthogonal to $\balpha$, as depicted in Figure 
\ref{kappagomega}(b).  The angle we are after is $\omega_U$, 
which is the angle between ${\bf c}$, the state vector of the 
$c$ quark, and the original tangent vector $\btau$.  This is 
easily seen in Figure \ref{kappagomega}(b), as can be checked
also by an explicit calculation using elementary differential 
geometry, to have half the value of the change in direction of 
the tangent vector itself, in the limit when the latter value is small.  
Hence, we have the result
\begin{equation}
\omega_U = \half \kappa_g \theta_{tc}.
\label{omegaUink}
\end{equation}
Similarly, of course,
\begin{equation}
\omega_D = \half \kappa_g \theta_{bs},
\label{omegaDink}
\end{equation}
although in this case $\kappa_g$ should in principle refer to the
geodesic curvature taken at $\mu = m_b$, not at $\mu = m_t$ as
in the equation above.  However, if we ignore this difference, which
is of order $\theta_{tb}$ compared with $\kappa_g$ and therefore
negligible to the order we are working, 
we obtain
\begin{equation}
\frac{V_{ub}}{V_{td}} \sim \frac{\sin \omega_U}{\sin \omega_D}
   \sim \frac{\sin \theta_{tc}}{\sin \theta_{bs}}.
\label{VubtoVtd}
\end{equation}
Taking the estimates obtained before in (\ref{thetatcbs}) for
these angles one then obtains the following estimate compared
to experiment,
\begin{equation}
\left[ \frac{V_{ub}}{V_{td}} \right]_{\rm est} \sim 0.56 \pm 0.01;
\quad
\left[ \frac{V_{ub}}{V_{td}} \right]_{\rm exp} = 0.40 \pm 0.03.
\label{VubtoVtde}
\end{equation}
This is as close an agreement as one can expect, since the starting
formulae (\ref{SFD2}) are correct only to order $\delta s \sim
\theta_{tb}$ and from (\ref{thetatb}) one would expect an error
in the matrix elements of the order of $\theta_{tb}^2 \sim 0.0017$,
which is not much smaller than the actual values of the matrix 
elements themselves.

So far one has input from experiment only the values of the mass 
ratios (\ref{thetatcbs}) and the fact that $\theta_{tb}$ is small.
One can go even further still to estimate the actual values of 
the corner elements by inputting in addition $\theta_{tb}$ from 
(\ref{thetatb}) to set the scale and, say, the Cabibbo angle 
$V_{us}$ from (\ref{VCKM}) to estimate the value of the geodesic 
curvature $\kappa_g$.  Indeed, using the formulae for $V_{us}$ in 
(\ref{Vckm2}), and for $\omega_U, \omega_D$ in (\ref{omegaUink}) 
and (\ref{omegaDink}), one easily obtains
\begin{equation}
\kappa_g \sim 3.0,
\label{kappagest}
\end{equation}
which when substituted back into (\ref{omegaUink}) and
(\ref{omegaDink}) will give
\begin{equation}
\omega_U \sim 0.128 \pm 0.004, \quad
\omega_D \sim 0.23 \pm 0.03.
\end{equation}
This then gives the following estimates for the actual values of 
the corner elements compared with experiment,
\begin{eqnarray}
 & & V^{\rm est}_{ub} \sim 0.0053\pm0.0002, \quad 
V^{\rm exp}_{ub} = 0.00347^{+0.00016}_{-0.00012};\nonumber\\
&  & V^{\rm est}_{td} \sim 0.0094^{+0.0014}_{-0.0011}, \quad
V^{\rm exp}_{td} = 0.00862^{+0.00026}_{-0.00020}.
\label{VubVtdest}
\end{eqnarray}
Again the agreement is about as good as can be expected, given 
the intrinsic errors in the starting Serret--Frenet--Darboux 
formulae of first order.

Now that one has reproduced by the above means the 
corner elements, one can proceed to give values
to all elements of the CKM matrix.  First from the estimated values of 
$\omega_U, \omega_D$ and of $\kappa_g$, one can evaluate the 
elements $V_{cb}$ and $V_{ts}$ as $(\cos \omega_U) \theta_{tb}$ 
and $(\cos \omega_D) \theta_{tb}$ by (\ref{Vckm2}).  For the other
elements we shall make use of unitarity, i.e.,
the condition that that every row or column should have norm 1,
rather than the formula (\ref{Vckm2}) accurate only to order 
$\theta_{tb}$, because this will give the diagonal elements to 
a better accuracy.  Indeed, even for the pair $V_{cb}$ and $V_{ts}$ we
could have used unitarity applied to the last 
row and column in (\ref{Vckm2}) because these satisfy unitarity 
already to order $\theta_{tb}^2$, and would have got identical results.
Then applying unitarity to 
the remaining rows and columns, one obtains
\begin{equation}
V_{\rm CKM}^{\rm out} = \left( \begin{array}{lll}
                0.97427 & {\it 0.2253} & 0.00530 \\
                0.2252 & 0.97346 & 0.0408 \\
                0.00943 & 0.0401 &{\it 0.999152} 
                \end{array} \right),
\label{CKMout}
\end{equation}
where the entries in italics are inputs, the rest being evaluated
as explained.  This compares very well with  (\ref{VCKM}) from
experiment.

One sees thus, that with the simple formula (\ref{Vckm2}) deduced
from rotation, one can go already quite some way towards explaining
the general features of the CKM matrix, in particular its corner
elements and their asymmetry about its diagonal.  In principle, the 
same arguments can be applied to the PMNS matrix for leptons as well, 
but there are a couple of serious reservations.  First, the expansion
parameter $\delta s$ is now given, in parallel to $\theta_{tb}$
above, by $U_{\tau 3}$ in (\ref{UPMNS}) as $\sin \theta_{\tau 3}
\sim 0.70$, which will take a lot of imagination to consider as
a small expansion parameter.  Secondly, on the physical masses of 
the neutrinos as given in experiment, the question arises whether 
they should be regarded as the Dirac masses which figure in the 
rotation formulae of (\ref{hiermass}) above, or as the masses 
obtained from these via some see-saw mechanism.  
This question matters in the estimate 
for the rotation angles $\theta_{\nu_3\nu_2}$ between $\nu_3$, the heaviest 
neutrino, and the next heaviest $\nu_2$, which is related to
the Dirac masses.  However, let us be 
cavalier here and boldly ignore the first 
reservation \footnote{This cavalier attitude is not really necessary, except here for the 
sake of simplicity and transparency, for the Serret-Frenet-Darboux 
formulae can readily be integrated by numerical methods given some
assumptions about the rotation trajectory.  An investigation along 
these lines to reproduce the whole mixing matrix and the mass spectrum of both quarks and leptons 
is near completion, and a report is under preparation 
\cite{compmec},
but the results obtained on the corner elements are 
not qualitatively different from those given below.}, while for the 
second, simply repeat the arguments for both the cases suggested.

Starting then from the experimental values of  $m_\tau  = 
1.777$, $m_\mu = 0.106$ in GeV [12], we first obtain
\begin{equation}
\sin \theta_{\tau \mu} = \sqrt{m_\mu/m_\tau} = 0.244,
\label{thetataumu}
\end{equation}
the experimental errors here being negligible.  Then taking
the experimental neutrino mass differences \cite{databook},
\begin{eqnarray}
& \Delta m_{32}^2 = (2.43 \pm 0.13) \times 10^{-3}\;{\rm eV}^2, \nonumber \\ 
& \Delta m_{21}^2 = (7.59 \pm 0.21) \times 10^{-5}\;{\rm eV}^2,
\label{numasses}
\end{eqnarray}
and assuming normal hierarchy along with a negligible $m_1$
we obtain:
\begin{equation}
m_3 = 0.050 \pm 0.001\;{\rm eV}, \ \ m_2 = 0.0087 \pm 0.0001\;{\rm eV}.
\label{yyyy}
\end{equation}
Hence for the Dirac case (D), we have straighforwardly:
\begin{equation}
\sin \theta^{\rm D}_{\nu_3\nu_2} = \sqrt{m_2/m_3} \sim 0.417 \pm 0.008.
\label{zzzz}
\end{equation}
For the see-saw case (ss), we take for the see-saw mechanism 
\cite{seesaw}  the simplest model, i.e., Type I quadratic 
see-saw, where 3 right-handed neutrinos are introduced and 
where their mass matrix is assumed to be proportional to the 
identity $3 \times 3$ matrix.  This then gives the physical 
masses of the 3 neutrino states $\nu_i$ as respectively $m_i 
= (m^{\rm D}_i)^2/m_R$, with $m^{\rm D}_i$ being their Dirac masses and 
$m_R$ the right-handed neutrino mass.  It follows therefore
that their Dirac masses $m^{\rm D}_i$ are proportional to the 
square-root of their physical masses $m_i$ as given above in
(\ref{yyyy}).  And since it is the Dirac mass which enters
in the rotation mechanism, we deduce that in the (ss) case:
\begin{widetext}
\begin{equation}
\sin\theta^{\rm ss}_{\nu_3\nu_2} = \sqrt{m^{\rm D}_2/m^{\rm D}_3} 
= \sqrt{\sqrt{m_2/m_3}}
   = \sqrt{\sin \theta^{\rm D}_{\nu_3\nu_2}} \sim 0.646 \pm 0.006.
\label{wwww}
\end{equation}
\end{widetext}
We note first that the error in this rotation angle induced 
by the experimental errors on neutrino masses is quite small,
for both (D) and (ss), as to be entirely negligible for the 
purpose we shall make use of it to deduce the results below.
Secondly, and more importantly, since for both cases (D and 
ss), $\sin \theta_{\tau \mu} < \sin \theta_{\nu_3\nu_2}$, it follows
that the corner elements of the PMNS matrix will have the same 
asymmetry as in the CKM matrix, i.e. $U_{e3} < U_{\tau 1}$, 
as seems to be implied by experiment in (2).  In fact, for 
the values adopted there for $\theta_{12}$ and 
$\theta_{23}$, this 
asymmetry will persist up to $\sin^2 2 \theta_{13}=0.23$ or to 
$\theta_{13} =  0.25$, i.e., way outside the experimental errors.

Secondly, accepting the result (\ref{omegaDink}) obtained above
even for this case, namely that the angles (\ref{wwww}) can 
still be considered as small enough for the arguments in Figure
\ref{kappagomega} to apply, so that for charged leptons and neutrinos
respectively we have:
\begin{equation}
\omega_U = \half \kappa_g \theta_{\tau \mu}, \ \ 
\omega_D = \half \kappa_g \theta_{\nu_3\nu_2},
\label{omegalepink}
\end{equation}
and assuming again that $\kappa_g$ is the same for $U$ and $D$,
we obtain
\begin{equation}
\frac{U_{e3}}{U_{\tau 1}} \sim \frac{\theta_{\tau \mu}}{\theta_{\nu_3\nu_2}} 
   \sim 0.6\; ({\rm D}), \  0.4\; ({\rm ss}),
\label{Ucornelratio}
\end{equation}
as compared with the value $0.47$ obtained for the 
matrix in (\ref{UPMNS}) above and agreeing with the noted asymmetry.  
Lastly, pushing it all the way, 
if we repeat the previous arguments to estimate the values of the
corner elements for the CKM matrix, inputting here in place of the
Cabibbo angle the solar neutrino angle $\sin^2(2 \theta_{12}) \sim 
0.861$ \cite{databook}, and in place of the element $V_{tb}$ the
element $U_{\tau 3}$ as given in (\ref{UPMNS}) obtained by unitarity 
from a maximal $U_{\mu 3}$ (the effect of the corner $U_{e3}$ being
negligible in this calculation for the accuracy needed), we obtain 
the estimates
\begin{equation}
U_{e3}^{\rm D} \sim 0.06; \ \ U_{e3}^{\,\rm ss} \sim 0.05.
\label{Ue3out}
\end{equation}
These estimates, though obviously very crude, are not ridiculous,
and maintain the above observation that the corner elements will
be small and be asymmetric about the diagonal to roughly the order
as seems indicated by experiment.  Notice in particular that the
corner elements here, just as in the CKM case before, have no
reason to vanish as it has in some symmetry schemes, unless, of 
course, the geodesic curvature happens to be exactly zero at that 
point of the rotation trajectory under consideration while being
nonzero both at $\mu=m_t$ and at $\mu=m_b$.  Besides, if $\kappa_g=0$,
then according to (\ref{Vckm2}) and (\ref{omegalepink}), the solar 
neutrino angle $U_{e2}$
would vanish also, in gross contradiction to experiment.

In the above study of both the CKM and PMNS matrices, we notice 
that in arriving at the conclusion that the corner elements are
small and have the right asymmetry, and differ from one another
even by about the right ratios, one has input from experiment 
only $m_c/m_t, m_s/m_b$ for CKM,  and $m_\mu/m_\tau, 
m_2/m_3$ for PMNS, mass ratios which have {\it a priori} 
nothing to do with the up-down mixing contained in the CKM and 
PMNS matrices, except via the R2M2 hypothesis.  Hence, that one 
has come to the right conclusions can justly be regarded as 
a nontrivial test of the rotation hypothesis.  
To a lesser extent, perhaps, even 
the rough estimates for the sizes of the corner elements can also be
regarded as such, for here one has input in addition 
from experiment only two other mixing matrix elements in each
case, namely $V_{tb},V_{us}$ for CKM and $U_{\tau 3}, U_{e2}$ 
for PMNS, which are not enough to determine via unitarity the
corner elements.

Finally, we come to the important question of the CP-violating
phase, which so far has been ignored.  A very interesting point
of the R2M2 hypothesis is that it also automatically provides an
explanation for this phase, and in quite an intriguing manner, by 
relating it to the theta-angle of topological origin in the QCD 
action, thereby even offering a solution to the old strong CP
problem.  This comes about because R2M2 has the very distinctive
property that the fermion mass matrix remains of rank one
throughout, hence having two zero eigenvalues at every scale.
These zero eigenstates, as is well-known \cite{Weinbergbook}, would
allow the strong theta-angle to be eliminated by a chiral 
transformation without making the mass matrix complex.  Yet,
because of the ``leakage mechanism'' due to rotation described 
above in (\ref{mc2}), none of the quarks need have a zero physical 
mass.  Since the mass matrix rotates with scale, however, the
chiral transformation, which is to be performed at every scale
in the direction of the normal vector $\bnu$ of the Darboux triad
at that scale, has to be scale-dependent.  Hence, its effects 
will get transmitted on to the state vectors of the various 
quarks by rotation, and further on to the CKM matrix also, where 
they will appear as a CP-violating phase.  The details of how 
this actually happens are explained in \cite{r2m2, ato2cps} to 
which the reader is referred.  However, whether a similar procedure 
applies also to 
leptons is at present unclear.

The beauty of the approximate formula (\ref{Vckm2}) is that it 
allows the above effect of the chiral transformation required 
for eliminating the theta-term to be given explicitly.  We first 
expand the $U$- and $D$-triads respectively in terms of the Darboux 
triads at $\mu = m_t$ and $\mu = m_b$.  Then we apply to each
the appropriate chiral transformation, each in the direction 
of the vector $\bnu$ at its own scale, giving for the relevant 
left-handed fields just a phase $\exp (-i \theta/2)$ to that 
component in the $\bnu$ direction,

\begin{widetext}
\begin{align}
(\bt,\bc,\bu) &= (\balpha,\,\cos \omega_U \btau +\sin 
\omega_U \bnu\,
e^{-i \theta/2},   \,\cos \omega_U \bnu\, e^{-i\theta/2}
\!-\sin \omega_U \btau)\nonumber,\\
(\bb,\bs,\bd)& = (\balpha', \,\cos \omega_D \btau' + 
\sin \omega_D
\bnu'\, e^{-i \theta/2},  \,\cos \omega_D \bnu' \,e^{-i  \theta/2}
\!-\sin \omega_D \btau').
\label{UDtriadsCP} 
\end{align}
\end{widetext}
The CKM matrix can then be evaluated with these two triads of 
state vectors as before by (\ref{Vckm}), but this will now 
contain a CP-violating phase.

To see what effect this phase so obtained will have, it would
be easiest to evaluate the Jarlskog invariant corresponding to
this matrix. Notice however that in the case of R2M2, where 
the orientation of the mass matrix is scale dependent, one cannot relate 
CP non-conservation directly to the commutator of the (hermitian) mass matrices, as 
it was originally proposed by C. Jarlskog in \cite{Jarlskog}. Instead, one has to rely solely on the unitary properties of the mixing matrix and work with the quartic rephasing invariants which are scale independent.
These invariants appeared in earlier works on CP non-conservation \cite{CP-nonconservation} without mentioning the mass matrix commutator properties.

Suppose we take in particular the minor of the
matrix at the bottom right, the elements of which are then given
by (\ref{UDtriadsCP}) to be
\begin{eqnarray}
V_{cb} & = & - \cos \omega_U \,\theta_{tb} \\
V_{ts} & = & \cos \omega_D \,\theta_{tb} \\
V_{tb} & = & 1\\
V_{cs} & = & \cos (\omega_D - \omega_U) - \sin (\omega_D - \omega_U)
\kappa_g \,\theta_{tb} \,\cos (\theta/2) \nonumber \\
&&{}+ i \sin (\omega_D + \omega_U)
\kappa_g \,\theta_{tb} \,\sin (\theta/2).
\label{CKMminorCP}
\end{eqnarray}
We then arrive at the following explicit formula for the Jarlskog
invariant \cite{Jarlskog}
\begin{eqnarray}
J &=& \Im \{V_{cs}\, V_{tb}\, V^*_{cb}\, V^*_{ts} \}\\
&=&  -\cos \omega_U \cos \omega_D \sin(\omega_D + \omega_U)\, \kappa_g
\,\theta_{tb}^3 \,\sin (\theta/2).
\label{JarlskogJ}
\end{eqnarray}
Though approximate, this is much more compact and amenable than 
that obtained before in \cite{r2m2, ato2cps}.  For example, putting
in the values given above for $\omega_U, \omega_D, \theta_{tb}$ and
$\kappa_g$, one easily obtains
\begin{equation}
|J\,| \sim 7.1 \times \sin (\theta/2) \times 10^{-5},
\label{estJ}
\end{equation}
which, for the strong CP angle $\theta$ of order unity, is of the 
order of the experimental value of $J \sim 2.9 \times 10^{-5}$ as 
given in \cite{databook}.  Alternatively, inputting the experimental
value, one arrives at the estimate of $|\theta| \sim 0.8$, which is
indeed of order unity.

In principle one can redo the whole analysis above taking
account of the CP-violating phase all through.  Indeed we have done
just that.  However, rather than including the theta-term right at the
beginning, we think it much more transparent to present our
results in the way we did, knowing full well that for the accuracy we
aim for at present, the effect of the CP phase will be small 
because of the smallness of $J$, and the absolute values of the 
CKM matrix elements displayed will not be  much affected.   
This is confirmed by our calculations.

\vspace{.5cm}

In summary, we conclude that the R2M2 (rotation) hypothesis gives
automatically small but nonzero corner elements to both the CKM 
and PMNS matrices with the right asymmetry and roughly the right 
magnitudes as observed in experiment.  This is, we believe, a 
nontrivial test for the hypothesis.  The merit of the small angle 
approximation used here is the utter 
simplicity and transparency of its derivations, with the minimum 
of assumptions on the rotation trajectory and without resorting 
to numerical methods.  Thus, though not giving as extensive and 
as accurate results, it is a valuable complement to the approach
by numerical integration of the Serret-Frenet-Darboux formulae
based on an explicit parametrization of the rotation
trajectory \cite{compmec}.

\end{document}